\newcommand{\CLASSINPUTtoptextmargin}{1.2in} 
\newcommand{\CLASSINPUTbottomtextmargin}{1.6in} 
\documentclass[conference,a4paper]{IEEEtran}
\IEEEoverridecommandlockouts
\usepackage{cite}
\usepackage{amsmath,amssymb,amsfonts}
\usepackage{algorithmic}
\usepackage[ruled,vlined]{algorithm2e}
\usepackage{graphicx}
\usepackage{textcomp}
\usepackage{bm}
\usepackage{pgfplots}
\usepackage{balance}
\usepackage{nicefrac}
\usepgfplotslibrary{groupplots}
\usepackage{bbm}
\usepackage{tikz}
\usetikzlibrary{arrows,automata,plotmarks,matrix,scopes,fit,calc,shapes,positioning,decorations,intersections,fit,backgrounds}
\usepackage{xcolor}
\newtheorem{example}{Example}
 
\def\BibTeX{{\rm B\kern-.05em{\sc i\kern-.025em b}\kern-.08em
    T\kern-.1667em\lower.7ex\hbox{E}\kern-.125emX}}

\DeclareMathOperator*{\unequals}{\lesseqgtr}
\begin{document}

\title{{Soft-Decision Decoding for LDPC Code-Based Quantitative Group Testing} \\
\thanks{\textcolor{black}{This work was partly funded by the German Research Foundation (DFG) under Grant Agreement No. WA 3907/7-1 and by the Swedish Research Council (VR) under grant 2020-03687.}}
}

\newcommand{\marvin}[1]{{\footnotesize  [\textbf{\textcolor{red}{#1}} \textcolor{red!60!black}{--mxh}]\normalsize}}
\newcommand{\alex}[1]{{\footnotesize  [\textbf{\textcolor{magenta}{#1}} \textcolor{red!60!black}{--agia}]\normalsize}}
\newcommand{\johan}[1]{{\footnotesize  [\textbf{\textcolor{orange!90!black}{#1}} \textcolor{red!60!black}{--jo}]\normalsize}}
\newcommand{\TODO}[1]{\textcolor{red}{TODO: #1}}
\newcommand{\bfx}{\bm{x}}
\newcommand{\bfw}{\bm{w}}
\newcommand{\bfc}{\bm{c}}
\newcommand{\bfb}{\bm{b}}
\newcommand{\bfe}{\bm{e}}
\newcommand{\bfX}{\bm{X}}
\newcommand{\bft}{\bm{t}}
\newcommand{\bfz}{\bm{z}}
\newcommand{\bfs}{\bm{s}}
\newcommand{\bfv}{\bm{v}}
\newcommand{\bfu}{\bm{u}}
\newcommand{\bfd}{\bm{d}}
\newcommand{\bfT}{\bm{T}}
\newcommand{\bfH}{\bm{H}}
\newcommand{\bfA}{\bm{A}}
\newcommand{\calP}{\mathcal{P}}
\newcommand{\calX}{\mathcal{X}}
\newcommand{\calR}{\mathcal{R}}
\newcommand{\calD}{\mathcal{D}}
\newcommand{\calM}{\mathcal{M}}
\newcommand{\calN}{\mathcal{N}}
\newcommand{\calT}{\mathcal{T}}
\newcommand{\msv}{\mathsf{v}}
\newcommand{\msc}{\mathsf{c}}
\newcommand{\hbfd}{\hat{\bm{d}}}
\newcommand{\hd}{\hat{d}}
\newcommand{\msFA}{\mathsf{FA}}
\newcommand{\msMD}{\mathsf{MD}}
\newcommand{\msD}{\mathsf{D}}
\newcommand{\rowspan}{\ensuremath{\mathsf{Sp}_{\text{r}}}}

\newcommand{\nm}{n_{\mathsf{m}}}
\newcommand{\transpose}{^\mathsf{T}}
\newcommand{\bfa}{\bm{a}}
\newcommand{\tbfd}{\tilde{\bm{d}}}
\newcommand{\td}{\tilde{d}}
\newcommand{\tbfs}{\tilde{\bm{s}}}
\newcommand{\ts}{\tilde{s}}

\newcommand{\APP}{\mathsf{APP}}
\newcommand{\calE}{\mathcal{E}}
\newcommand{\dv}{d_\mathsf{v}}
\newcommand{\dc}{d_\mathsf{c}}

\definecolor{darkblue}{rgb}{0.07843,0.16863,0.54902}
\definecolor{darkgreen}{rgb}{0,0.49804,0}%
\definecolor{brown}{rgb}{0.85098, 0.32941, 0.10196}%

\newcommand{\temp}{\textcolor{red}}
\newcommand{\newtext}{\textcolor{black}}

\newcommand{\FL}{FL}

\author{\IEEEauthorblockN{Marvin Xhemrishi}
\IEEEauthorblockA{
\textit{Technical University of Munich}\\
Munich, Germany \\
\texttt{marvin.xhemrishi@tum.de}}
\and
\IEEEauthorblockN{ Johan \"Ostman}
\IEEEauthorblockA{
\textit{AI Sweden}\\
Gothenburg, Sweden \\
\texttt{johan.ostman@ai.se}}
\and
\IEEEauthorblockN{Alexandre Graell i Amat}
\IEEEauthorblockA{
\textit{Chalmers University of Technology}\\
Gothenburg, Sweden \\
\texttt{graell@chalmers.se}}
}

\maketitle

\begin{abstract}
We consider the problem of identifying defective items in a population with non-adaptive quantitative group testing. For this scenario, Mashauri \emph{et al.} recently proposed a low-density parity-check (LDPC) code-based quantitative group testing scheme with a 
hard-decision decoding approach (akin to peeling decoding). This scheme outperforms generalized LDPC code-based quantitative group testing schemes in terms of the misdetection rate. In this work, we propose a belief-propagation-based decoder for quantitative group testing with LDPC codes, where the messages being passed are purely soft. Through extensive simulations, we show that the proposed soft-information decoder outperforms the hard-decision decoder Mashauri \emph{et al.}. 
\end{abstract}


\section{Introduction}

Group testing encompasses a family of test schemes with the aim to identify items affected by some particular condition, usually referred to as \emph{defective} items (e.g., individuals infected by a virus), within a large population of $n$ items (e.g., all individuals). 
Dating back to the Second World War, group testing was pioneered by Dorfman~\cite{Dor43} to facilitate the identification of 
syphilis among soldiers in the US Army at low cost.

The primary objective of  group testing is to minimize the number of tests required to identify the defective items within the population.  The key idea is that, if the number of defective items is significantly smaller than $n$, then negative tests on pools of items can spare many individual tests. 
Following this principle, items are grouped into overlapping groups, and tests are performed on each group. 
Typically, binary tests are considered~\cite{Dor43}, where a positive test implies that at least one defective item participates in the corresponding group. Conversely, a negative test implies that all items in the group are non-defective.
After testing, the binary outputs serve as the input to a decoding algorithm that infers the status\textemdash  i.e., defective or non-defective\textemdash of each item in the population. In general, the inference has some probability of error associated with it, referred to as \emph{misdetection} rate. 

Since its introduction, group testing has emerged as an important and powerful tool for solving problems across several fields including biology, computer science, and data science~\cite{Ald19}. 
For example, during the COVID-19 pandemic, many public health institutions relied on group testing to identify patients infected by the virus~\cite{Gol20}. Group testing may also be used in situations where the goal is not to reduce the number of tests. For example, in~\cite{Xhe23}, group testing is used to privately identify malicious clients in federated learning with secure aggregation. 

While conventional group testing considers only binary test results, in quantitative group testing~\cite{Geb19} the test outcome is (ideally) equal to the exact number of defective items in the group, i.e., the test is akin to an \emph{adder} channel. \newtext{Similar to group testing, the quantitative variant has real-world applications, like in biology~\cite{Cao14}.} Leveraging the additional information provided by quantitative tests, the number of tests can be significantly reduced compared to group testing based on binary tests~\cite[Sec. 5.9]{Ald19}.


\textit{Related work:} Group testing is a well-researched area with connections to several fields, e.g., information theory~\cite{Ald19} and error-correction codes~\cite{Bar17,Lee19,Wad13,Avi18}. 
In~\cite{Ald19}, a plethora of fundamental results on the required number of tests are presented for several flavors of group testing.  
Moreover, these fundamental results are used to benchmark practical decoding algorithms for group testing schemes. 
An optimal decoding strategy for (noisy) group testing, based on the well-known forward-backward decoding algorithm~\cite{Bah74}, was presented in~\cite{Liv21}. 
However, due to its exponential complexity in the number of tests, the optimal decoder is not practical for large population sizes. For quantitative group testing, the complexity of the forward-backward decoding algorithm  is even higher. 
In~\cite{Kar19,Kar19-2}, the authors proposed a scheme for non-adaptive noiseless quantitative group testing based on generalized low-density parity-check (LDPC) codes with $w$-error correcting BCH codes as component codes. However, the strongest codes (large $w$) do not perform well with iterative decoding and the best performance is achieved for $w=2$\cite{Kar19,Kar19-2}. Recently, the authors in~\cite{Mas23} proposed a quantitative noiseless group testing scheme based on LDPC codes, together with  a corresponding hard-decision decoding algorithm, that outperforms the schemes in~\cite{Kar19,Kar19-2} in terms of misdetection rate. 
The scheme in \cite{Mas23} is considered state-of-the-art sparse-graph code-based quantitative group testing. \newtext{The authors in~\cite{Fei20} adopt techniques from compressed sensing~\cite{Don06} to solve the quantitative group testing problem. However, the proposed algorithms require the knowledge of the number of defective items in the population, which is in general not known.}

\textit{Contribution:} In this work, we adopt the quantitative group testing scheme based on  LDPC codes proposed in~\cite{Mas23}. Our novelty resides in a new iterative decoding approach for quantitative group testing that exchanges soft information. The decoder is inspired by the well-known \emph{belief propagation} decoding for LDPC codes, but utilizes different updates for the constraint nodes. Through simulation results, we show that 
the proposed decoder significantly outperforms the  hard-decision decoding approach in~\cite{Mas23} in terms of misdetection rate.  


\textit{Organization:} In Section~\ref{sec:prelim}, we define our notation. 
Section~\ref{sec:sys_model} describes the system model for quantitative group testing. 
In Section~\ref{sec:BP}, we present the soft-information decoder for quantitative group testing and its components. 
In Section~\ref{sec:num_results}, we present the performance of the proposed decoding approach obtained by simulation. \newtext{Section~\ref{sec:fut_work} highlights future work and Section~\ref{sec:conclusion} concludes the paper.}

\section{Notation}
\label{sec:prelim}


We use lowercase bold letters, e.g., $\bfx$, to denote row vectors. The $i$-th element of vector $\bfx$ is denoted by $x_i$. To represent matrices, we use uppercase bold letters, e.g., $\bfX$. 
Random variables representing scalars are denoted by uppercase letters, such as $X$. The probability mass function (PMF) of a random variable $X$ is denoted as $\Pr_{X}(x)$, where $x$ is a realization. We use calligraphic letters, such as $\calX$,  to denote sets with cardinality denoted by $\lvert \calX \lvert$. $\mathbb{F}_q$, with $q$ being a prime, denotes a finite field. For an integer $x$, we use the notation $[x]$ to denote the set of all positive integers less than or equal to $x$, i.e., $[x] = \{1, 2, \dots, x\}$. We use the symbol $\mathbbm{1}\{\cdot\}$ to denote the indicator function. 
  The Hamming weight of a vector $\bfx$ is  denoted by wt$(\bfx)$.



\section{System Model}
\label{sec:sys_model}

We consider a population of $n$ items  represented by a binary vector $\bfd =(d_1, d_2, \dots, d_n) \in \mathbb{F}_2^n$, where $d_i = 1$ if item $i$ is defective and $d_i = 0$ if it is not. We refer to $\bfd$  as the \emph{defective vector}, which is unknown.  

We consider a \emph{probabilistic} model for the status of the items, where $D_1, D_2, \dots, D_n$ are independently and identically distributed (i.i.d.) random variables following a Bernoulli distribution as
\begin{align*}
    {\Pr}_{D_i}(b) = \begin{cases}
        1 - \delta & \text{if}\;\;b = 0 \\
        \delta & \text{if}\;\; b=1
    \end{cases}\,, \quad i \in [n]\,.
\end{align*}
Adopting the terminology of group testing, we refer to $\delta$ as the \emph{prevalence}.

\begin{figure}[t]
    \centering
    \begin{tikzpicture}
\tikzstyle{mynode}=[draw,circle,minimum width=0.5cm,inner sep = 0cm, fill = gray!90!white, label distance=0.1cm];
\tikzstyle{CN}=[draw,rectangle,minimum height=0.5cm,minimum width = 0.5cm, fill = gray!30!white, inner sep = 0cm];

\matrix (m) [column sep=6.5mm, row sep=2cm, ampersand replacement=\&]
{
\node{}; 
\node[] (lv1) {\footnotesize $\msv_1$}; \&
\node[] (lv2) {\footnotesize $\msv_2$}; \&
\node[] (lv3) {\footnotesize $\msv_3$}; \&
\node[] (lv4) {\footnotesize $\msv_4$}; \&
\node[] (lv5) {\footnotesize $\msv_5$}; \&
\node[] (lv6) {\footnotesize $\msv_6$}; \&
\node[] (lv7) {\footnotesize $\msv_7$}; \&
\node[] (lv8) {\footnotesize $\msv_8$}; \\[-2cm]
\node{}; 
\node[mynode] (v1) {}; \&
\node[mynode] (v2) {}; \&
\node[mynode] (v3) {}; \&
\node[mynode] (v4) {}; \&
\node[mynode] (v5) {}; \&
\node[mynode] (v6) {}; \&
\node[mynode] (v7) {}; \&
\node[mynode] (v8) {}; \\
\node{}; \&
\node[CN] (c1) {}; \& \& %
\node[CN] (c2) {}; \&\&
\node[CN] (c3) {}; \& \&
\node[CN] (c4) {}; \\[-2cm]
\node{}; \&
\node[] (lc1) {\footnotesize $\msc_1$}; \& \&
\node[] (lc2) {\footnotesize $\msc_2$}; \&\&
\node[] (lc3) {\footnotesize $\msc_3$}; \& \&
\node[] (lc4) {\footnotesize $\msc_4$}; \\[]
};
\draw[solid, black, ultra thick] (v1) -- (c1);
\draw[solid, black, ultra thick] (v2) -- (c1);
\draw[solid, black, ultra thick] (v6) -- (c1);
\draw[solid, black, ultra thick] (v8) -- (c1);
\draw[solid, blue, ultra thick] (v1) -- (c2);
\draw[solid, blue, ultra thick] (v4) -- (c2);
\draw[solid, blue, ultra thick] (v5) -- (c2);
\draw[solid, blue, ultra thick] (v6) -- (c2);
\draw[solid, red, ultra thick] (v2) -- (c3);
\draw[solid, red, ultra thick] (v3) -- (c3);
\draw[solid, red, ultra thick] (v4) -- (c3);
\draw[solid, red, ultra thick] (v7) -- (c3);
\draw[solid, orange, ultra thick] (v3) -- (c4);
\draw[solid, orange, ultra thick] (v5) -- (c4);
\draw[solid, orange, ultra thick] (v7) -- (c4);
\draw[solid, orange, ultra thick] (v8) -- (c4);

\end{tikzpicture}
    \caption{Bipartite graph representation of the assignment matrix $\bfA$ in~\eqref{eq:A}. The circles denote the variable nodes (VN), while the squares represent the constraint nodes (CN). The connections of the items to their respective group are color-coded. }
    \label{fig:bipartite_graph}
\end{figure}
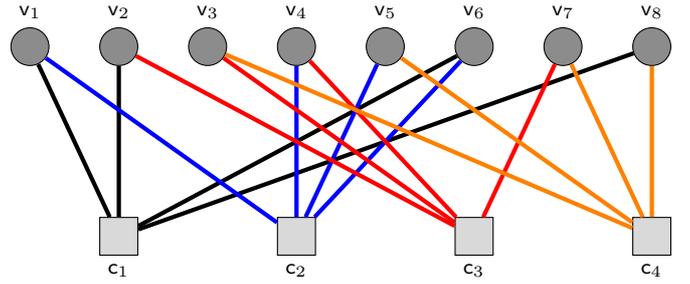

The main goal of a group testing scheme is to infer the defective vector $\bfd$ using $r$ tests. More precisely,  items are grouped into $r$ overlapping groups (also referred to as pools) $\calP_1, \calP_2, \dots, \calP_r$,
and a test is applied to each pool. We denote by $s_i$ the result of the test applied to pool $\calP_i$, and collect the test results for all pools in a  vector $\bfs = (s_1, s_2, \dots, s_r)$. The assignment of items to pools can be represented by an $r\times n$ binary adjacency matrix $\bfA=(a_{i,j})$, where $a_{i,j} = 1$ if item $j$ participates in pool $i$ and $a_{i,j} = 0$ if it does not.
We refer to $\bfA$ as the \emph{assignment} matrix. Note that  pool $\calP_i$, $i \in [r]$, is defined by the set
\begin{align*}
    \calP_i = \{j\lvert   a_{i, j} = 1, \, j \in [n]\}\,.
\end{align*}
The assignment matrix $\bfA$ establishes a link between group testing and error-correcting codes. 
In particular, the parity-check matrix of an error-correcting code can be used as the assignment matrix $\bfA$~\cite{Bar17}. 
The rate of the quantitative group testing scheme, denoted by $R$, is defined as the ratio\footnote{Note that the definition is slightly different from the rate used in error-correcting codes.} between the number of tests over the size of the population, i.e., $R = \frac{r}{n}$.


In this work, as in \cite{Kar19,Mas23}, we consider non-adaptive noiseless quantitative group testing, where the result of test $i$, $s_i$, yields the exact number of defective items in pool $\calP_i$. Hence, $s_i:\calP_i\rightarrow \{0,1,\ldots,|\calP_i|\}$, 
\begin{align*}
    s_i = \sum_{j \in \calP_i} d_j = \sum_{j = 1}^n d_j a_{i, j} 
\end{align*}
and
\begin{equation}
   \label{eq:syndrome}
   \bfs = \bfd\cdot\bfA^{\transpose} \,.
\end{equation}
Based on $\bfs$ and the assignment matrix $\bfA$, the goal of the quantitative group testing scheme is to infer the defective vector $\bfd$ through an inference (decoding) algorithm, $\hat{\bfd} = \text{Dec}(\bfs, \bfA)$.

\begin{example}[Assignment matrix]
\label{ex:A}
Consider the assignment matrix  
\begin{equation}
    \label{eq:A}
    \bfA = \begin{pmatrix}
        1 & 1 & 0 & 0 & 0 & 1 & 0 & 1 \\
        1 & 0 & 0 & 1 & 1 & 1 & 0 & 0 \\
        0 & 1 & 1 & 1 & 0 & 0 & 1 & 0 \\
        0 & 0 & 1 & 0 & 1 & 0 & 1 & 1
    \end{pmatrix}
\end{equation}
of regular row weight $4$ and column weight $2$, corresponding to a scenario with $n=8$ items. This matrix instructs the pooling into $r=4$ groups as $\calP_1 = \{1,2,6,8\}, \calP_2 = \{1,4,5,6\}, \calP_3 = \{2,3,4,7\}$, and $\calP_4 = \{3,5,7,8\}$.
\end{example}

The assignment matrix and its imposed grouping can be graphically represented by a bipartite graph consisting of  $n$ variable nodes (VNs) and $r$ constraint nodes (CNs), corresponding to the $n$ items and $r$ tests, respectively. An edge connects CN $\msc_i$ with VN $\msv_j$ if item $j$ participates in pool $\calP_i$. 
  The corresponding bipartite graph of the assignment matrix $\bfA$ in~\eqref{eq:A} is illustrated in Fig.~\ref{fig:bipartite_graph}. It is a $(\dv,\dc)=(2,4)$ regular graph, where $\dv$ is the VN degree and $\dc$ the CN degree.


We will use the terminology $\calN(x)$ to denote the set of neighbors of node $x$ in the graph, i.e., the set of nodes adjacent to node $x$. It holds that the neighbors of CN $\msc_i$ and VN $\msv_i$ are 
\begin{align*}
    \calN(\msc_i) &= \{ \msv_j \;\lvert\; j \in \calP_i \}, \quad i \in [r]\\
    \calN(\msv_j) &= \{ \msc_i \;\lvert\; j \in \calP_i, i\in [r]\}, \quad j \in [n]\,.
\end{align*}
For a regular graph, $\lvert \calN(\msc_i) \lvert = \dc, \; i \in [r]$, and $\lvert \calN(\msv_j) \lvert = \dv, \; j \in [n]$ and the rate can be written as $R = \nicefrac{\dv}{\dc}$.

\section{Belief-propagation Decoding for\\ Quantitative Group Testing}
\label{sec:BP}

In this section, we introduce the main contribution of this paper, the belief-propagation (BP) decoding algorithm for quantitative group testing. 


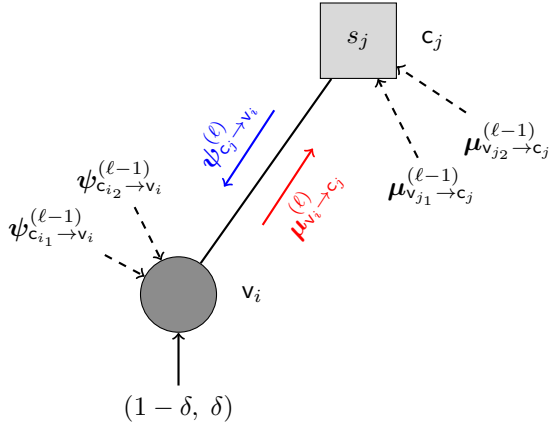
\begin{figure}[t]
    \centering
    \begin{tikzpicture}
\tikzstyle{VN}=[draw,circle,minimum width=1cm,inner sep = 0cm, fill = gray!90!white];
\tikzstyle{CN}=[draw,rectangle,minimum height=1cm,minimum width = 1cm, fill = gray!30!white, inner sep = 0cm];

\node[CN] (CN) at (0,-1) {$s_j$};
\node[VN, below left = 2.5cm and 1.5cm of CN] (VN) {}; 
\node[below = 0.7cm of VN] (prev) {$\left(1-\delta, \;\delta\right)$};
\node[left = 2mm of VN] (v_i) {};
\node[right = 2mm of VN] (v_iii) {$\msv_i$};
\node[above left = 4mm and 0mm of v_i] (vizaji) {$\bm{\psi}^{(\ell-1)}_{\msc_{i_1} \rightarrow \msv_i}$};
\node[above left = 8mm and -3mm of VN] (viza) {$\bm{\psi}^{(\ell-1)}_{\msc_{i_2} \rightarrow \msv_i}$};
\node[right = 2mm of CN] (c_j) {$\msc_j$};
\node[below right = 7mm and 0mm of c_j] (VNjj) {$\bm{\mu}^{(\ell -1 )}_{\msv_{j_2} \rightarrow \msc_j}$};
\node[below right = 13mm and -10mm of c_j] (VNj) {$\bm{\mu}^{(\ell-1)}_{\msv_{j_1} \rightarrow \msc_j}$};

\draw[thick, solid] (CN) -- (VN);
\coordinate (mid) at ($(CN)!0.5!(VN)$);
\draw[thick,red,->, shorten < = 1.1cm, shorten >= 1.1cm] (VN.east) -- node[midway, sloped, below] {\textcolor{red}{$\bm{\mu}^{(\ell)}_{\msv_{i} \rightarrow \msc_j}$}}(CN.south);
\draw[thick, ->] (prev) -- (VN);
\draw[thick,blue,->, shorten < = 1.1cm, shorten >= 1.1cm] (CN.west) -- node[midway, sloped, above] {\textcolor{blue}{$\bm{\psi}^{(\ell)}_{\msc_{j} \rightarrow \msv_i}$}}(VN.north);
\draw[thick, dashed, ->] (vizaji) -- (VN);
\draw[thick, dashed, ->] (viza) -- (VN);
\draw[thick, dashed, ->] (VNj) -- (CN);
\draw[thick, dashed, ->] (VNjj) -- (CN);

\end{tikzpicture}
    \caption{Visualization of the variable and constraint node updates shown in~\eqref{eq:VN_update} and~\eqref{eq:CN_update}, respectively. The circle represents a variable node $\msv_i$, while the square represents a constraint node $\msc_j$. Note that the constraint node $\msc_j$ has the test outcome of the $j$-th pool $s_j$ as a constraint the incoming messages should fulfill. }
    \label{fig:CN_VN_update}
\end{figure}

The decoding algorithm has to infer the defective vector $\hat{\bfd}$ given the test outcome vector $\bfs$, prevalence $\delta$, and assignment matrix $\bfA$. Implementing a MAP decoder  to compute the a posteriori probabilities $\Pr(D_i|\bfs)$ as in~\cite{Liv21} has complexity $\mathcal{O}(2^n)$, which is infeasible for large assignment matrices $\bfA$. The BP algorithm attempts to approximate the a posteriori probabilities operating on the graph of the quantitative group testing scheme. This approximation is obtained using an iterative message-passing approach between variable and constraint nodes. A visualization of a message-passing between a constraint and variable node is depicted in Fig.~\ref{fig:CN_VN_update}. The (estimated) a posteriori probabilities can be used as soft-information to infer the defective vector $\hat{\bfd}$.


Let  $\bm{\mu}^{(\ell)}_{\msv \rightarrow \msc} = \left({\mu}^{(\ell)}_{\msv \rightarrow \msc}(0), {\mu}^{(\ell)}_{\msv \rightarrow \msc}(1)\right)$
be the message from VN $\msv$ to CN $\msc$ at iteration $\ell$, corresponding to the probability that the item is non-defective (${\mu}^{(\ell)}_{\msv \rightarrow \msc}(0)$) and defective (${\mu}^{(\ell)}_{\msv \rightarrow \msc}(1)$).
Similarly, let $\bm{\psi}^{(\ell)}_{\msv \rightarrow \msc} = \left({\psi}^{(\ell)}_{\msv \rightarrow \msc}(0), {\psi}^{(\ell)}_{\msv \rightarrow \msc}(1)\right)$ be the message from CN $\msc$ to VN $\msv$ at iteration $\ell$, corresponding to the belief from the CN that the item is non-defective and defective, respectively.  The pseudo-code for the proposed BP decoder for quantitative group testing is shown in Algorithm~\ref{alg:BP_decoder}. In the following, we define the VN and CN updates.



 


\RestyleAlgo{ruled}
\begin{algorithm}[!t]
    \caption{Belief-propagation decoding for quantitative group testing}\label{alg:BP_decoder}
      \SetKw{Input}{Input:}
    \SetKw{Output}{Output:}
    \SetKw{Initialization}{Init:}
    \Input{$\bfs$, $\delta$, $\mathcal{N}\!\left(\msc_i\right)$ for $i \in [r]$, $\mathcal{N}\!\left(\msv_i\right)$ for $i \in [n]$, $L$.}\\[.5ex]
    \Output{$\hat{\bfd}$.}\\[.5ex]
    \Initialization{$\bm{\psi}^{(0)}_{\msc_i \rightarrow \msv_j} = (0.5, 0.5)$ for $i \in [r]$, $\msv_j \in \mathcal{N}\left(\msc_i\right)$} \\[.5ex]
    \begin{algorithmic}[1]
    \FOR{$\ell = 1,2, \dots, L$}
    \STATE /* VN Update */
    \FOR{$i = 1,2, \dots, n$}
    \STATE Compute $\bm{\mu}^{(\ell)}_{\msv_i \rightarrow \msc_j}$ for $\msc_j \in \mathcal{N}\!\left(\msv_i\right)$ as in~\eqref{eq:VN_update}
    \ENDFOR
    \STATE /*CN Update*/
     \FOR{$j = 1,2, \dots, r$}
    \STATE Compute $\bm{\psi}^{(\ell)}_{\msc_j \rightarrow \msv_i}$ for $\msv_i \in \mathcal{N}\!\left(\msc_j\right)$ as in~\eqref{eq:CN_update}
    \ENDFOR
    \STATE /* \emph{A posteriori} computation and decision */
    \FOR{$i = 1,2, \dots, n$}
    \STATE{$\Pr\{D_i = b \lvert \bfs\} \approx \Pr_{D_i}(b) \prod_{\msc_j \in \mathcal{N}\left(\msv_i \right)}\psi^{(\ell)}_{\msc_j \rightarrow \msv_i} (b)$} \\
    \STATE $\Pr\{D_i = 0 \lvert \bfs\} \unequals_{\hat{d}_i = 0}^{\hat{d}_i = 1} \Pr\{D_i = 1 \lvert \bfs\}$
    \ENDFOR
    \STATE /* Stopping criterion */
    \IF{$\hat{\bfd}\bfA == \bfs$}
    \STATE{\bf break}
    \ENDIF
    \ENDFOR
    \end{algorithmic}
\end{algorithm}

\subsection{Variable Node Update}
\label{subsec:VN_update}


The message from a VN $\msv_i$ to a CN $\msc_j$ is simply the product of all incoming message to VN $\msv_i$, except the one on the edge $(\msv_i,\msc_j)$, and the prior on the value $b$:
\begin{equation}
    \label{eq:VN_update}
    \mu_{\msv_i\rightarrow \msc_j}^{(\ell)}(b) = {\Pr}_{D_i}(b) \cdot\prod_{\msc'\in \mathcal{N}(\msv_i)\setminus \msc_j}\psi_{\msc'\rightarrow \msv_i}^{(\ell - 1)}(b)
\end{equation}
for $\ell \in \mathbb{Z}^+, b \in \mathbb{F}_2, i \in [n]$ and $\msc_j \in \mathcal{N}\!\left(\msv_i\right)$. 

\subsection{Constraint Node Update}

As illustrated in Fig.~\ref{fig:CN_VN_update}, the test outcome of the $j$-th pool is associated with  CN $\msc_j$ for $j \in [r]$. 
A test outcome $s_j$ means that there are exactly $s_j$ defective items among the $\mathcal{N}\!\left(\msc_j\right)$ members of the $j$-th pool. 
Therefore, the constraint of  CN $\msc$ with  associated test outcome $s$ can be written as 
\begin{equation}
    \label{eq:constraint}
    \mathbbm{1}\left\{ \sum_{\msv \in \calN\left(\msc \right)} b_{\msv} = s \right\}\,,
\end{equation}
where $b_{\msv} \in \{0,1\}$ is the value of VN $\msv$.\footnote{This constraint is somewhat similar to that defined in~\cite{Ros18}, which considers LDPC codes for counter braids.}
 
The CN update can then be written as
\begin{align}
    \label{eq:CN_update}
    \psi^{(\ell)}_{\msc_i\rightarrow \msv_j}(b) &= \sum_{\msv'\in \mathcal{N}(\msc_i)\setminus \msv_j}
    \sum_{b_{\msv'} \in \mathbb{F}_2 } \;\mathbbm{1}\left\{\sum_{\msv' \in \mathcal{N}(\msc_i) \setminus \msv_j} b_{\msv'}  = s_i-b\right\}  \\
    &\qquad\qquad\cdot\!\prod_{\msv' \in \mathcal{N}(\msc_i)\setminus \msv_j} \mu^{(\ell -1)}_{\msv' \rightarrow \msc_i}(b_{\msv'})  \nonumber
\end{align}
for $\ell \in \mathbb{Z}^+,  i \in [r]$ and $\msv_j \in \mathcal{N}\!\left(\msc_i\right)$. The main idea behind the CN update in~\eqref{eq:CN_update} is that if $b=0$, then $s_i$ defective items should be in $\calN\!\left(\msc_i\right) \setminus \msv_j$. Hence, the CN update considers $\binom{\lvert \calN\left(\msc_i\right) \lvert - 1}{s_i}$ possibilities that pass the constraint in~\eqref{eq:constraint} and multiplies the incoming messages from its neighboring VNs. Similarly, $b=1$  implies that $s_i - 1$ defective items are in $\calN\!\left(\msc_i\right) \setminus \msv_j$ and it considers $\binom{\lvert \calN\!\left(\msc_i\right) \lvert - 1}{s_i - 1}$ possibilities that pass the constraint.  

\begin{example}[Constraint node update]
\label{ex:CN_update}
We will take the example depicted in Fig.~\ref{fig:CN_VN_update} to further clarify the CN update. As illustrated in the figure, the CN $\msc_j$ sends the belief $\bm{\psi}^{(\ell)}$ to the VN $\msv_i$. Let us assume that the test outcome is $s_j = 1$. Then,  using~\eqref{eq:CN_update}, for $b=0$, the defective node should be one of the two neighbors, namely $\msv_{j_1}$ or $\msv_{j_2}$. Hence one can write
$${\psi}_{\msc_j \rightarrow \msv_i}^{(\ell)}(0) = \sum_{a \in \mathbb{F}_2}{\mu}^{(\ell-1)}_{\msv_{j_1}\rightarrow\msc_j}(a)\cdot{\mu}^{(\ell-1)}_{\msv_{j_2}\rightarrow\msc_j}(1-a)\,.$$ On the other hand, for the message $b=1$, following~\eqref{eq:CN_update}, the indicator function requires that  
only the incoming messages representing $0$ will be considered (since $s_j-1 =0$). Hence, one can write $${\psi}_{\msc_j \rightarrow \msv_i}^{(\ell)}(1) = {\mu}^{(\ell-1)}_{\msv_{j_1}\rightarrow\msc_j}(0)\cdot{\mu}^{(\ell-1)}_{\msv_{j_2}\rightarrow\msc_j}(0)\,.$$
\end{example}
The CN update defined in~\eqref{eq:CN_update} generalizes and encompasses the CN update rule from the peeling decoder  in~\cite{Mas23}, since 
\begin{itemize}
    \item If $s_i = 0$, then~\eqref{eq:CN_update} implies that $\left(\psi^{(\ell)}_{\msc_i\rightarrow \msv_j}(0), \psi^{(\ell)}_{\msc_i\rightarrow \msv_j}(1)\right) = (1, 0)$ ({after normalizing}), which is the same as declaring the $j$-th node as non-defective ($\hat{d}_j = 0$). 
    \item If $s_i = \lvert \calN\left(\msc_i\right) \lvert - r_i^{(\ell)}$, where $r_i^{(\ell)}$ is the number of resolved non-defective members of $\calN\left(\msc_i\right)$ in the $\ell$-th iteration, then~\eqref{eq:CN_update} implies that $\left(\psi^{(\ell)}_{\msc_i\rightarrow \msv_j}(0), \psi^{(\ell)}_{\msc_i\rightarrow \msv_j}(1)\right) = (0, 1)$ (for $\msv_j$ representing an unresolved member of $\calN\left(\msc_i\right)$). This is equivalent to declaring the $j$-th node as defective ($\hat{d}_j = 1$). 
\end{itemize}


\subsection{Computation Complexity}
\label{subsec:Comp_complexity}
{A soft-information decoder generally increases the decoding complexity compared to a hard-decision decoder (such as the peeling decoder in~\cite{Mas23}). Our proposed decoder is no exception to that rule. \newtext{As described in Algorithm~\ref{alg:BP_decoder}, the decoding algorithm is iterative with $L$ iterations. Each iteration involves $n\cdot \dv$ operations for the VNs and $r$ CN updates. Clearly, the CN update requires more computations than the VN one. From~\eqref{eq:CN_update}, we conclude that the number of computations for the CN update  is $\binom{\dc -1}{s} + \binom{\dc - 1}{s -1}$, where $s$ is the test outcome. Hence, we can upper bound the CN update complexity with $\mathcal{O}\left(2^{\dc - 1}\right)$. Since the CN update is the most computationally heavy part of the decoding algorithm, we state that the complexity of the proposed soft-decision decoding algorithm is $\mathcal{O}\left(2^{\dc - 1}Lr\right)$. For small values of $\dc$, the decoding algorithm is feasible, while for large number of members in the pool the decoder becomes practically infeasible.} 

\section{Numerical results}
\label{sec:num_results}
In this section, we present the performance of our proposed decoder in a quantitative group testing scenario obtained by numerical simulations. 
As a benchmark, we consider the peeling decoder introduced in~\cite{Mas23}, which outperforms other sparse-graph codes-based schemes such as the schemes in \cite{Kar19,Kar19-2} based on generalized LDPC codes. 
As a figure of merit, we use the misdetection rate for the performance, which aligns with the group testing literature. 
The misdetection rate is defined as 
\begin{equation}
    \label{eq:MD}
    P_{\msMD} = \dfrac{1}{n} \sum_{i=1}^{n} \Pr \left\{\hat{D}_i = 0\lvert D_i = 1\right\}\,.
\end{equation}

We numerically estimate $P_{\msMD}$ for the peeling decoder in~\cite{Mas23} and for the proposed decoder using numerical simulations. 
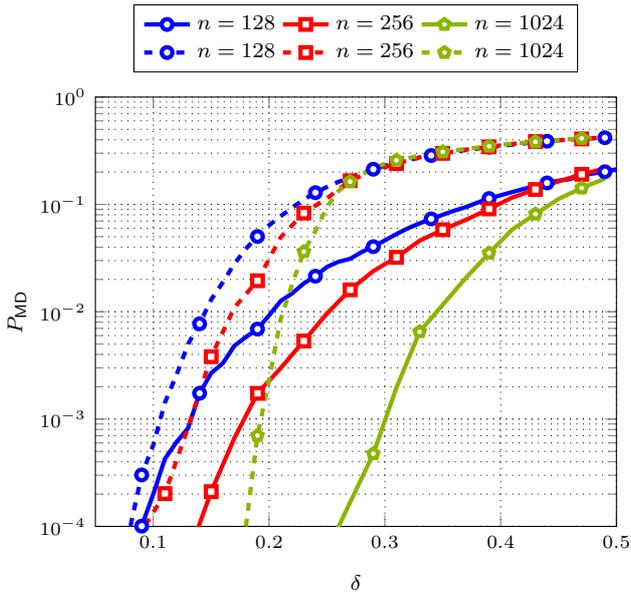
\begin{figure}[t]
    \centering
    \definecolor{applegreen}{rgb}{0.55, 0.71, 0.0}
\begin{tikzpicture}

\begin{groupplot}[ 
group style={
	group size=1 by 1,
	horizontal sep=2cm, 
},
]
\nextgroupplot
[
grid = both, 
grid style={dotted,draw=black!90},
tick label style={/pgf/number format/fixed, font =\scriptsize},
xmode = linear, 
ymode = log, 
ymax = 1, 
ymin = 1e-4, 
xmax = 0.5, 
xmin = 0.05,
xlabel = \footnotesize $\delta$,
ylabel = \footnotesize $P_{\msMD}$,
ylabel style={yshift=-2mm},
legend style =
{
	fill=white,
        legend columns = 3,
	draw=black,
	anchor=center,
	legend cell align=center,
	align=center
},
legend to name=acc_legend
]
\addplot[blue, ultra thick, solid, mark=*, mark options = {fill = white}, mark repeat =5, mark size=2pt] table [x index = {0}, y index={1}, col sep=comma]
{./plots/csv_files/QGT-128_dc-6.csv}; 
\addlegendentry{\footnotesize $n = 128$}

\addplot[red, ultra thick, solid, mark=square*, mark options = {fill = white}, mark repeat =2, mark size=2pt] table [x index = {0}, y index={1}, col sep=comma]
{./plots/csv_files/QGT-256_dc-6.csv}; 
\addlegendentry{\footnotesize $n = 256$}

\addplot[applegreen, ultra thick, solid, mark=pentagon*, mark options = {fill = white}, mark repeat =2, mark size=2pt] table [x index = {0}, y index={1}, col sep=comma]
{./plots/csv_files/BP_QGT-1024_dc-6.csv}; 
\addlegendentry{\footnotesize $n = 1024$}

\addplot[blue, ultra thick, dashed, mark=*, mark repeat =5, mark options={fill=white, solid}, mark size=2pt] table [x index = {0}, y index={2}, col sep=comma]
{./plots/csv_files/QGT-128_dc-6.csv}; 
\addlegendentry{\footnotesize $n = 128$}

\addplot[red, ultra thick, dashed, mark=square*, mark repeat =2, mark options={fill=white, solid}, mark size=2pt] table [x index = {2}, y index={3}, col sep=comma]
{./plots/csv_files/QGT-256_dc-6.csv}; 
\addlegendentry{\footnotesize $n = 256$}

\addplot[applegreen, ultra thick, dashed, mark=pentagon*, mark options = {fill = white,solid}, mark repeat =4, mark size=2pt] table [x index = {0}, y index={1}, col sep=comma]
{./plots/csv_files/Peel_QGT-1024_dc-6.csv}; 
\addlegendentry{\footnotesize $n = 1024$}




\end{groupplot}
\node[above] at ([yshift=2mm]group c1r1.north)
{\pgfplotslegendfromname{acc_legend}};

\end{tikzpicture}
    \caption{We present the performance in terms of $P_{\msMD}$ versus $\delta$ for a regular graph with $\dv =3$ and $\dc =6$, that yields a rate $R = 0.5$. The dashed line shows the performance of the peeling decoder in~\cite{Mas23}, while the solid lines show the performance of the proposed soft decoder. The results are shown for the short-length regime $n \in \{128, 256, 1024\}$. The lines are color and marker-coded, such that the same color and marker are used for the same length $n$ and only the line style (dashed or solid) determines the decoder.}
    \label{fig:short_length}
\end{figure}
In the simulations, we use $L = 100$ iterations for the BP decoder\footnote{From preliminary results, we observed that using slightly fewer iterations, yielded similar performances. However, if the number of iterations was drastically reduced, then the performance of the BP decoder degraded significantly.}. 
Similar to~\cite{Mas23}, we  only consider assignment matrices based on regular LDPC codes,  but we note that the proposed decoder is compatible also with irregular LDPC codes.  
The defective vector $\bfd$ is randomly picked, with each entry being i.i.d. following a Bernoulli distribution with probability of success $\delta$. 
As we are interested in applying the decoder to a real-world problem, we investigate the performance of the proposed decoder in the short and moderate blocklength regime and under relatively high values of $\delta$ (corresponding, e.g., to the scenario of identifying malicious clients in federated learning~\cite{Xhe23}).

We consider an assignment matrix $\bfA$ corresponding to a $(\dv,\dc)=(3,6)$ regular graph and length $n \in \{128, 256, 1024, 2048, 8192, 16384\}$. In Fig.~\ref{fig:short_length}, we plot the misdetection rate, $P_{\msMD}$, of the proposed decoder (solid lines) and the peeling decoder in~\cite{Mas23} (dashed lines) in the short-length regime ($n < 2000$) as a function of the prevalence $\delta$. The performance is displayed in terms of $P_{\msMD}$ as defined in~\eqref{eq:MD} versus the prevalence $\delta$. {As expected, the performance improves with increasing $n$; for a fixed $(\dv,\dc)$, with increasing $n$ the graph gets closer to becoming cycle-free, allowing for a good performance from iterative decoders. }

\begin{figure}[t]
    \centering
    \definecolor{applegreen}{rgb}{0.55, 0.71, 0.0}
\begin{tikzpicture}

\begin{groupplot}[ 
group style={
	group size=1 by 1,
	horizontal sep=2cm, 
},
]
\nextgroupplot
[
grid = both, 
grid style={dotted,draw=black!90},
tick label style={/pgf/number format/fixed, font =\scriptsize},
xmode = linear, 
ymode = log, 
ymax = 1, 
ymin = 1e-4, 
xmax = 0.50, 
xmin = 0.15,
xlabel = \footnotesize $\delta$,
ylabel = \footnotesize $P_{\msMD}$,
ylabel style={yshift=-2mm},
legend style =
{
	fill=white,
        legend columns = 3,
	draw=black,
	anchor=center,
	legend cell align=center,
	align=center
},
legend to name=acc_legend_long
]
\addplot[blue, ultra thick, solid, mark=*, mark options = {fill = white}, mark repeat =2, mark size=2pt] table [x index = {0}, y index={1}, col sep=comma]
{./plots/csv_files/BP_QGT-2048_dc-6.csv}; 
\addlegendentry{\footnotesize $n = 2048$}

\addplot[red, ultra thick, solid, mark=square*, mark options = {fill = white}, mark repeat =2, mark size=2pt] table [x index = {0}, y index={1}, col sep=comma]
{./plots/csv_files/BP_QGT-8192_dc-6.csv}; 
\addlegendentry{\footnotesize $n = 8192$}

\addplot[applegreen, ultra thick, solid, mark=pentagon*, mark options = {fill = white}, mark repeat =2, mark size=2pt] table [x index = {0}, y index={1}, col sep=comma]
{./plots/csv_files/BP_QGT-16384_dc-6.csv}; 
\addlegendentry{\footnotesize $n = 16384$}

\addplot[blue, ultra thick, dashed, mark=*, mark options = {fill = white, solid}, mark repeat =5, mark size=2pt] table [x index = {0}, y index={1}, col sep=comma]
{./plots/csv_files/Peel_QGT-2048_dc-6.csv}; 
\addlegendentry{\footnotesize $n = 2048$}

\addplot[red, ultra thick, dashed, mark=square*, mark options = {fill = white, solid}, mark repeat =2, mark size=2pt] table [x index = {0}, y index={1}, col sep=comma]
{./plots/csv_files/Peel_QGT-8192_dc-6.csv}; 
\addlegendentry{\footnotesize $n = 8192$}

\addplot[applegreen, ultra thick, dashed, mark=pentagon*, mark options = {fill = white, solid}, mark repeat =2, mark size=2pt] table [x index = {0}, y index={1}, col sep=comma]
{./plots/csv_files/Peel_QGT-16384_dc-6.csv}; 
\addlegendentry{\footnotesize $n = 16384$}



\end{groupplot}
\node[above] at ([yshift=2mm]group c1r1.north)
{\pgfplotslegendfromname{acc_legend_long}};

\end{tikzpicture}
    \caption{In this plot, we show the performance of our proposed decoder (solid lines) and compare it to the peeling decoder in~\cite{Mas23} (dashed lines) for a regular graph $\dv = 3,\dc =6$ for moderate lengths $n \in \{2048, 8192, 16384\}$. The plot shows that for all choices of $n$, the misdetection probability $P_{\msMD}$ achieved by the proposed decoder is much lower compared to the one by peeling.}
    \label{fig:mod_length}
\end{figure}
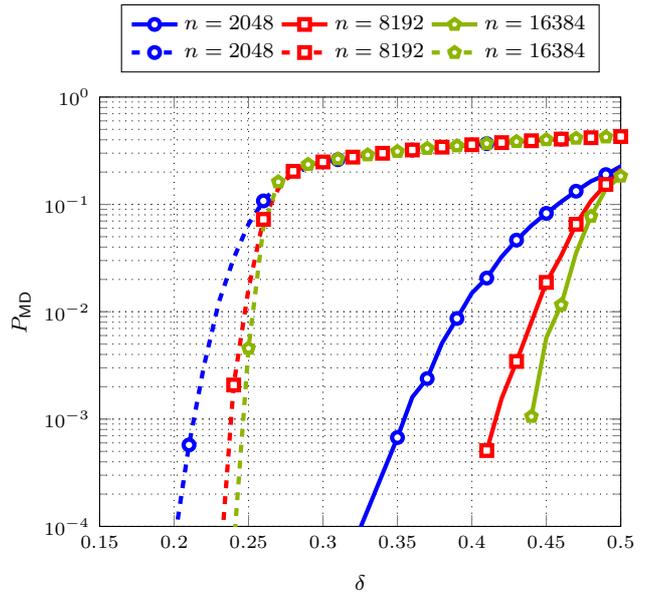
We observe that, for all considered $n$, the proposed soft-decision decoder outperforms the peeling decoder in~\cite{Mas23}. For a target misdetection rate $P_{\msMD} \approx 10^{-3}$, our decoder has a {gain}\footnote{With a slight abuse of terminology, by gain in the prevalence, we mean how much higher is the prevalence for a fixed misdetection probability.} of $ \approx 0.02, 0.04$ and $0.1$ in terms of $\delta$ for $n = 128, 256$ and $1024$, respectively.  

In Fig.~\ref{fig:mod_length}, we show the performance of our decoder for larger graphs, namely for $n \in \{2048, 8192, 16384\}$. In the moderate-length regime $( 2000 < n < 20000)$, we observe even higher {gains} in terms of $\delta$ for all the  considered lengths. For a target misdetection rate $P_{\msMD} \approx 10^{-3}$, one can see that the BP decoder has a {gain} of $\approx 0.14-0.18$ compared to the peeling decoder~\cite{Mas23}. In general, for all regular graphs with $\dv =3$ and $\dc = 6$, the proposed decoder achieves better performance compared to the peeling decoder~\cite{Mas23} and the {gain} in terms of the prevalence $\delta$ increases with the population size. 

\begin{figure}[t]
    \centering
    \definecolor{applegreen}{rgb}{0.55, 0.71, 0.0}
\begin{tikzpicture}

\begin{groupplot}[ 
group style={
	group size=1 by 1,
	horizontal sep=2cm, 
},
]
\nextgroupplot
[
grid = both, 
grid style={dotted,draw=black!90},
tick label style={/pgf/number format/fixed, font =\scriptsize},
xmode = linear, 
ymode = log, 
ymax = 1, 
ymin = 1e-3, 
xmax = 0.5, 
xmin = 0.05,
xlabel = \footnotesize $\delta$,
ylabel = \footnotesize $P_{\msMD}$,
ylabel style={yshift=-2mm},
legend style =
{
	fill=white,
        legend columns = 2,
	draw=black,
	anchor=center,
	legend cell align=center,
	align=center
},
legend to name=acc_legend
]
\addplot[blue, ultra thick, solid, mark=*, mark options = {fill = white}, mark repeat =1, mark size=2pt] table [x index = {0}, y index={1}, col sep=comma]
{./plots/csv_files/BP_QGT-4095_dc-9.csv}; 
\addlegendentry{\footnotesize $n = 4095$}

\addplot[red, ultra thick, solid, mark=square*, mark options = {fill = white}, mark repeat =1, mark size=2pt] table [x index = {0}, y index={1}, col sep=comma]
{./plots/csv_files/BP_QGT-16380_dc-9.csv}; 
\addlegendentry{\footnotesize $n = 16380$}

\addplot[blue, ultra thick, dashed, mark=*, mark options = {fill = white, solid}, mark repeat =5, mark size=2pt] table [x index = {0}, y index={1}, col sep=comma]
{./plots/csv_files/Peel_QGT-4095_dc-9.csv}; 
\addlegendentry{\footnotesize $n = 4095$}

\addplot[red, ultra thick, dashed, mark=square*, mark options = {fill = white, solid}, mark repeat =5, mark size=2pt] table [x index = {0}, y index={1}, col sep=comma]
{./plots/csv_files/Peel_QGT-16380_dc-9.csv}; 
\addlegendentry{\footnotesize $n = 16380$}

\end{groupplot}
\node[above] at ([yshift=2mm]group c1r1.north)
{\pgfplotslegendfromname{acc_legend}};

\end{tikzpicture}
    \caption{This plot shows the results as $P_{\msMD}$ versus $\delta$ for a regular graph with $\dv =3$ and $\dc = 9$ for $n \in \{4095, 16380\}$. The rate of the scheme defined by this regular graph is $R=\nicefrac{1}{3}$. The performance of our proposed decoder is shown in solid lines, while the performance of the peeling decoder~\cite{Mas23} is shown in dashed lines. For both choices of $n$, the proposed decoder outperforms the peeling decoder. }
    \label{fig:dc_9}
\end{figure}
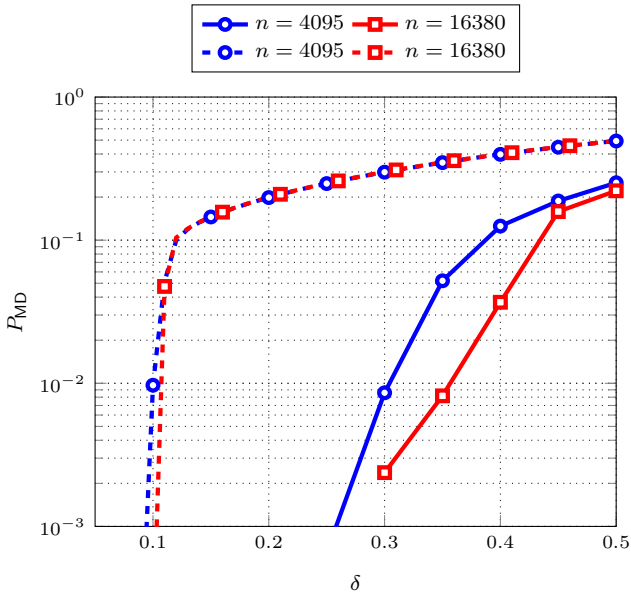

In Fig.~\ref{fig:dc_9}, we show the performance of the proposed decoder for a regular graph with $(\dv,\dc)=(3,9)$ and  size $n \in \{4095, 16380\}$. We observe that for a target misdetection rate $P_{\msMD} = 10^{-3}$, the decoder has a {gain} of around $\approx 0.16$ in the prevalence $\delta$ compared to the peeling decoder~\cite{Mas23}. 

Our simulation results show that for the two considered regular graphs, the proposed decoder clearly outperforms the peeling decoder in~\cite{Mas23}, for short-length and moderate-length regime.  

\section{Future Work}\label{sec:fut_work}

In contrast to the schemes in \cite{Kar19,Kar19-2,Mas23}, which, due to their hard-decision decoding nature,  cannot deal with noisy tests directly, the proposed soft-decision decoder allows to tackling noisy group testing by incorporating proper soft information into the decoder. In noisy quantitative group testing, the test outcome vector $\bft$ and the syndrome $\bfs$ (true test values) are not necessarily the same, and usually, the noisiness of the tests is modeled according to a probability distribution $Q(\bft \lvert \bfs)$. The implication of the noise is that the constraint equation in the constraint node is not necessarily correct. We are currently investigating the belief propagation decoder for noisy quantitative tests. Noisy tests have not been considered in the realm of sparse-graph code-based group testing schemes.

\newtext{On the other hand, the proposed decoder has an exponential computation complexity in the check node degree, which makes it infeasible for very large check node degrees. We leave for future work the derivation of suboptimal check nodes rules that are computationally less demanding.}

\section{Conclusion}
\label{sec:conclusion}

In this work, we presented a novel soft-decision iterative decoder for quantitative group testing based on low-density parity-check codes. We presented a belief propagation decoder and we derived the appropriate variable node and constraint node update rules tailored to a noiseless non-adaptive quantitative group testing scheme. Simulation results show that the proposed decoder significantly outperforms the state-of-the-art for the two regular graph choices. The {gain} in the prevalence $\delta$ for a target misdetection rate $P_{\msMD} = 10^{-3}$ varies from $0.02$ for short-length graphs to $0.17$ for moderate-length graphs.


\bibliographystyle{ieeetr}
\bibliography{references}

\end{document}